\documentclass[a4paper, reqno, 12pt]{amsart}
\pdfoutput=1

\usepackage[left=2.5cm, right=2.5cm, top=3cm, bottom=3cm]{geometry}

\usepackage{amsaddr}
\usepackage{amssymb}
\usepackage{amsfonts}
\usepackage{mathrsfs}
\usepackage[T1]{fontenc}
\usepackage[USenglish]{babel}
\usepackage{varioref}
\usepackage{multirow}
\usepackage{array,multicol}
\usepackage{bigdelim}
\usepackage{bigstrut}
\usepackage[font=small, labelfont=bf]{caption}
\usepackage[list=true, labelformat=brace, font=small, labelfont=bf]{subcaption}
\usepackage{graphicx}
\usepackage{csquotes}

\usepackage[breaklinks=true
]{hyperref}
\hypersetup{pdftitle={F.Nill: Endemic Oscillations}}
\usepackage[backend=biber,style=authoryear-comp]{biblatex}

\usepackage[inline]{enumitem}
\setlist{noitemsep}

\usepackage{url}

\def\customauthor{\empty}
\def\customdate{\empty}
\let\oldauthor\author
\renewcommand{\author}[1]{\def\customauthor{#1}}
\renewcommand{\date}[1]{\def\customdate{#1}}
\oldauthor{\customauthor\\\customdate}

\theoremstyle{definition}
\newtheorem{definition}{Definition}[section]

\theoremstyle{plain}
\newtheorem{theorem}[definition]{Theorem}

\newtheorem{corollary}[definition]{Corollary}

\theoremstyle{remark}

\numberwithin{equation}{section}

\newcommand{\ncm}{\newcommand}
\ncm{\rncm}{\renewcommand}

\ncm{\lb}[1]{\label{#1}}

\rncm{\sec}{\setc{0}\section}
\ncm{\bsn}{\bigskip\noindent}
\ncm{\beq}{\begin{equation}}
\ncm{\beqnon}{\begin{equation*}}
\ncm{\eeq}{\end{equation}}
\ncm{\eeqnon}{\end{equation*}}
\ncm{\bea}{\begin{eqnarray}}
\ncm{\beanon}{\begin{eqnarray*}}
\ncm{\eea}{\end{eqnarray}}
\ncm{\eeanon}{\end{eqnarray*}}

\ncm{\ba}{\begin{array}}
\ncm{\ea}{\end{array}}

\ncm{\fns}{\footnotesize}

\DeclareMathOperator\age{age}

\newcommand{\RR}{\ensuremath{\mathbb{R}}}
\ncm{\NN}{\ensuremath{\mathbb{N}}}
\ncm{\ZZ}{\ensuremath{\mathbb{Z}}}
\ncm{\GG}{\ensuremath{\mathbb{G}}}
\newcommand{\PP}{\ensuremath{{\mathcal P}}}
\ncm{\PPe}{\PP_\epsilon}

\newcommand{\bp}{{ \bf p}}
\ncm{\pt}{\bp_\tau}

\newcommand{\G}{\ensuremath{{\mathcal G}}}

\newcommand{\X}{\ensuremath{{\mathcal X}}}

\newcommand{\T}{\ensuremath{{\mathcal T}}}
\rncm{\P}{\ensuremath{{\mathcal P}}}
\rncm{\S}{\ensuremath{{\mathcal S}}}

\newcommand{\R}{\ensuremath{{\mathcal R}}}

\renewcommand{\L}{\ensuremath{{\mathcal L}}}
\ncm{\Labc}{\L_{a,b,c}}
\newcommand{\I}{\ensuremath{{\mathcal I}}}

\newcommand{\M}{\ensuremath{{\mathcal M}}}

\rncm{\O}{\mathcal{O}}

\ncm{\Me}{\M_\epsilon}

\ncm{\Pph}{P_{\mathrm{phys}}}
\ncm{\Tph}{\T_{\mathrm{phys}}}
\ncm{\TABC}{T_{ABC}}
\ncm{\Tabc}{\TABC^{\leq 1}}
\ncm{\Dph}{D_{\mathrm{phys}}}
\ncm{\bDph}{\bar{D}_{\mathrm{phys}}}
\ncm{\yph}{y_{\mathrm{phys}}}
\ncm{\Padm}{P_{\mathrm{adm}}}
\ncm{\Pc}{\P_{\mathrm{cut}}}
\ncm{\Tosc}{T_{\mathrm{osc}}}
\ncm{\Timm}{T_{\mathrm{imm}}}
\ncm{\Tinf}{T_{\mathrm{inf}}}
\ncm{\Thalf}{T_{\mathrm{half}}}
\ncm{\rvac}{r_{\mathrm{vac}}}

\newcommand{\si}{\sigma}
\newcommand{\ep}{\epsilon}

\newcommand{\om}{\omega}

\ncm{\OP}{\Omega_{\PP,\ep}}
\ncm{\oG}{\omega_\G}
\ncm{\ome}{\omega_\ep}
\ncm{\phit}{\varphi_\tau}
\ncm{\p}{\psi}
\newcommand{\al}{\alpha}

\newcommand{\be}{\beta}

\newcommand{\ga}{\gamma}
\newcommand{\gd}{\gamma_\delta}

\ncm{\Aal}{A_\alpha}
\ncm{\Bal}{B_\alpha}
\ncm{\sal}{\sigma_\alpha}

\rncm{\k}{\kappa}
\ncm{\an}{a_\nu}
\def\cros{\,\raise1.9pt\hbox{$\scriptscriptstyle  > $}\!
          \raise1.5pt\hbox{$\scriptstyle\triangleleft$}\,}
\def\>cros{\cros}
\def\<cros{\,\raise1.5pt\hbox{$\scriptstyle\triangleright$}\!
           \raise1.9pt\hbox{$\scriptscriptstyle < $}\,}

\ncm{\dR}{\partial_R}
\ncm{\dS}{\partial_S}
\ncm{\dI}{\partial_I}
\ncm{\dD}{\partial_D}
\ncm{\dN}{\partial_N}
\ncm{\dM}{\partial_M}
\ncm{\dX}{\partial_X}
\ncm{\dq}{\partial_q}
\ncm{\dx}{\partial_x}
\ncm{\dy}{\partial_y}
\ncm{\parH}{\partial H}
\ncm{\parHe}{\partial H_{\epsilon}}
\ncm{\parq}{\partial q}
\ncm{\parp}{\partial p}

\ncm{\rto}{\rightarrow}
\ncm{\mto}{\longmapsto}
\ncm{\lto}{\longrightarrow}
\ncm{\Lto}{\Longrightarrow}
\ncm{\LRA}{\Leftrightarrow}
\ncm{\LLRA}{\Longleftrightarrow}
\ncm{\LRa}{\Leftrightarrow}
\ncm{\LLRa}{\Longleftrightarrow}

\ncm{\tOP}{\tilde{\Omega}_{\PP,\ep}}
\ncm{\toe}{\tilde{\omega}_\ep}
\ncm{\tHe}{\tilde{H}_{\epsilon}}
\ncm{\tH}{\tilde{H}}
\ncm{\tV}{\tilde{V}}
\ncm{\tK}{\tilde{K}}
\ncm{\tE}{\tilde{E}}
\ncm{\rt}{\tilde{r}_0}
\ncm{\tr}{\tilde{r}_0}
\ncm{\tga}{\tilde{\gamma}}
\ncm{\Nt}{\tilde{N}}
\ncm{\tHPe}{\tilde{H}_{\PP,\epsilon}}
\ncm{\tre}{\tilde{\rho}_{\epsilon}}
\ncm{\tq}{\tilde{q}}
\ncm{\tp}{\tilde{p}}
\ncm{\etq}{e^{\tilde{q}}}
\ncm{\etp}{e^{\tilde{p}}}
\ncm{\ttau}{\tilde{\tau}}

\ncm{\hH}{\hat{H}}
\ncm{\hV}{\hat{V}}
\ncm{\hK}{\hat{K}}
\ncm{\hKs}{\hat{K}_\sigma}
\ncm{\hHs}{\hat{H}_\sigma}

\ncm{\s}{\mathsf{s}}
\ncm{\g}{\mathsf{g}}
\ncm{\h}{\mathsf{h}}
\ncm{\HIG}{H_\alpha}
\ncm{\Hal}{H_\alpha}
\ncm{\oIG}{\omega_\alpha}
\ncm{\HPLV}{H_{\mathrm {pLV}}}
\ncm{\omPLV}{\om_{\mathrm {pLV}}}
\ncm{\Hreg}{H^{\mathrm {reg}}}
\ncm{\omreg}{\om^{\mathrm {reg}}}
\ncm{\HPLVreg}{H_{\mathrm {pLV}}^{\mathrm {reg}}}
\ncm{\omPLVreg}{\omega_{\mathrm {pLV}}^{\mathrm{reg}}}
\ncm{\Halreg}{H_{\alpha}^{\mathrm {reg}}}
\ncm{\omalreg}{\omega_{\alpha}^{\mathrm {reg}}}
\ncm{\Lt}{L_\tau}
\ncm{\Lmin}{L_{\min}}
\ncm{\dLt}{\dot{L}_\tau}
\ncm{\rv}{a_\mathrm{vac}}

\newcommand{\Del}{\Delta}

\newcommand{\Eqref}[1]{Eq. \eqref{#1}}

\ncm{\ulim}[1]{\underset{#1}{\lim}}
\ncm{\secref}[1]{Section \ref{#1}}
\ncm{\figref}[1]{Fig. \ref{#1}}

\newcommand{\minus}{\scalebox{0.75}[1.0]{$-$}}
\newcommand{\inv}{^{\minus 1}}
\ncm{\vsir}{V_{SIR}}
\ncm{\Vsir}{V_{SIR}}
\ncm{\hsir}{H_{SIR}}

\ncm{\zit}[1]{\autocite{#1}}
\ncm{\GHS}{\textsc{HGS }}
\ncm{\Upot}{$U\!$-potential}
\ncm{\QUpot}{quasi-\Upot}
\ncm{\UE}{{V^E}}
\ncm{\VE}{{V^E}}
\ncm{\alpm}{\upsilon_\pm}
\ncm{\alp}{\upsilon_+}
\ncm{\alm}{\upsilon_-}
\ncm{\vpm}{\upsilon_\pm}
\ncm{\xpm}{x_\pm}
\ncm{\qpm}{q_\pm}

\ncm{\vp}{\upsilon_+}
\ncm{\vm}{\upsilon_-}
\ncm{\vO}{v_{\O}}
\ncm{\vN}{v_N}
\ncm{\vt}{v_\tau}
\ncm{\ut}{u_\tau}
\ncm{\apm}{a_\pm}
\ncm{\bpm}{b_\pm}
\ncm{\upm}{u_\pm}
\ncm{\qp}{q_+}
\ncm{\qc}{q_c}
\ncm{\up}{u_+}
\ncm{\xp}{x_+}
\ncm{\xc}{x_c}
\ncm{\epm}{\varepsilon_\pm}
\ncm{\fpm}{f_\pm}
\ncm{\Apm}{A_\pm}
\ncm{\Bpm}{B_\pm}
\ncm{\Dpm}{D_\pm}
\ncm{\Dp}{\vp}
\ncm{\Dc}{\Delta_c}

\ncm{\Spm}{S_\pm}
\ncm{\rSpm}{\rho S_\pm}
\ncm{\thpm}{\theta_\pm}

\ncm{\ntg}{\notag\\}

\ncm{\Ss}{S_{\textsl{sample}}}
\ncm{\Is}{I_{\textsl{sample}}}
\ncm{\Zs}{Z_{\textsl{sample}}}
\ncm{\Es}{E_{\textsl{sample}}}
\ncm{\Ns}{N_{\textsl{sample}}}
\ncm{\rhos}{\rho_{\textsl{sample}}}
\ncm{\gs}{\gamma_{\textsl{sample}}}
\ncm{\Zrge}{Z_{\rho,\gamma,E}}
\ncm{\Zmax}{Z_{\max}}
\ncm{\el}{e^{\lambda}}
\ncm{\Ve}{V_{\epsilon}}
\ncm{\He}{H_{\epsilon}}

\ncm{\HPe}{H_{\PP,\epsilon}}
\ncm{\Emax}{E_{\max}}
\ncm{\re}{\rho_{\epsilon}}
\ncm{\qe}{q_{\epsilon}}
\ncm{\expe}{\exp_{\epsilon}}
\ncm{\lne}{\ln_{\epsilon}}
\ncm{\Vpme}{V_{\pm,\ep}}
\ncm{\Vpe}{V_{+,\ep}}
\ncm{\Vme}{V_{-,\ep}}
\ncm{\qpme}{q_{\pm,\ep}}
\ncm{\qpe}{q_{+,\ep}}
\ncm{\qme}{q_{-,\ep}}
\ncm{\xpme}{x_{\pm,\ep}}
\ncm{\xpe}{x_{+,\ep}}
\ncm{\xme}{x_{-,\ep}}
\ncm{\qG}{q_\G}
\ncm{\pG}{p_\G}
\ncm{\ys}{y_2^*}
\ncm{\xs}{x_1^*}
\ncm{\vs}{v_2^*}
\ncm{\us}{u_1^*}
\ncm{\fl}{\varphi_\tau}
\ncm{\Gas}{\Gamma_\sigma}
\ncm{\tmp}{\age}
\ncm{\tImp}{\tau_{\I,\max}}

\begin{document}

\title[Endemic Oscillations for Omicron]
{Endemic Oscillations for SARS-CoV-2 Omicron -
\\
A SIRS model analysis
}
\author{Florian Nill}

\address{Department of Physics, Free University Berlin, Arnimallee 14, 14195 Berlin, Germany.}
\date{Revised May 30, 2023}

\email{florian.nill@fu-berlin.de}



\begin{abstract}
The SIRS model with constant vaccination and immunity 
waning rates is well known to show a transition from a  
disease-free to an endemic equilibrium as the basic 
reproduction number $r_0$ is raised above threshold. It 
is shown that this model maps to  Hethcote's classic 
endemic model originally published in 1973. In this way 
one obtains unifying formulas for a whole class of 
models showing endemic bifurcation.  In particular, if 
the vaccination rate is smaller than the recovery rate 
and $r_-<r_0<r_+$ for certain upper and lower bounds 
$r_\pm$, then trajectories spiral into the endemic 
equilibrium via damped infection waves. Latest data of 
the SARS-CoV-2 Omicron  variant suggest that according 
to this simplified model continuous vaccination programs 
will not be capable to escape the oscillating endemic 
phase. However, in view of the strong damping factors 
predicted by the model, in reality 
these oscillations will certainly be overruled by 
time-dependent contact behaviors.
\end{abstract}

\maketitle

%

\begin{center}
{\em All models are wrong, but some are useful}
[George E.P.Box]
\end{center}

\section{Introduction}

According to actual estimates the basic reproduction number $r_0$ for the Delta- and Omicron-variants of Covid-19 ranges between 
$r_{0,\mathrm{Delta}}\approx 5	-9$
and $r_{0,\mathrm{Omicron}}\approx 7-14$.%
\footnote{Determining the basic reproduction number empirically is not an exact science. There are many methods and model dependent definitions and empirical data are volatile. Authors mostly refer to effective reproduction numbers and data also depend on regional authority measures. So in this paper I will only rely on ranges of magnitude. For overviews based on U.S. CDC-reports see
\\
\url{https://www.npr.org/sections/goatsandsoda/2021/08/11/1026190062/covid-delta-variant-transmission-cdc-chickenpox/}\,,
\url{https://health-desk.org/articles/how-contagious-is-the-delta-variant-compared-to-other-infectious-diseases}\,,
\\
\url{https://www.cdc.gov/coronavirus/2019-ncov/variants/about-variants.html/}\,. Also see the CA PHO-report
\url{https://www.publichealthontario.ca/-/media/documents/ncov/covid-wwksf/2022/01/wwksf-omicron-communicability.pdf/}\,.
}
Experts therefore seem to agree, that Omicron will completely take over and cause Covid-19 to run into an endemic scenario no matter how strong contact preventing and/or vaccination measures are enforced. For an epidemiological discussion of the transition to endemicity for Covid-19 see \autocites{AntiaHalloran}. As explained by the authors, when approaching the endemic limit prevalence typically does not decrease monotonically, but there are several waves of infection. 
These are affected by non-pharmaceutical interventions, increased transmissibility due to virus evolution and of course intrinsic seasonality in transmission. The purpose of the present paper is to analyze when and to what extend damped oscillations would also be predicted by a classic autonomous (i.e. with static coefficients)  endemic SIR-type model.

\begin{figure}[ht!]
\centering
\includegraphics[width=0.8\textwidth]
	{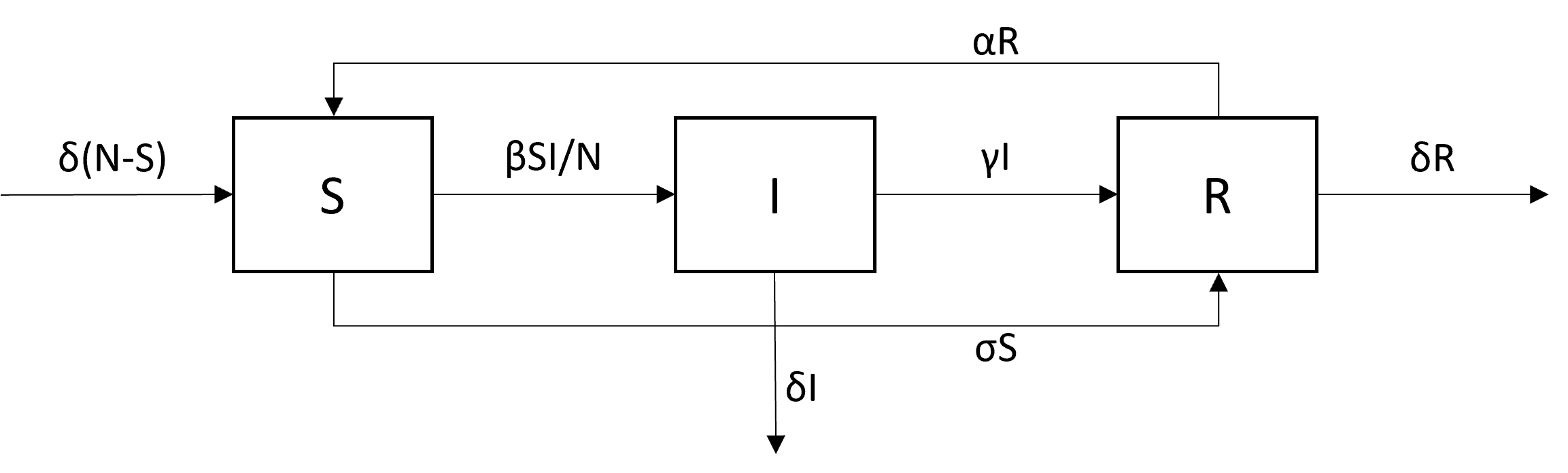}
	\caption{Flow diagram of a SIRS model with effective contact rate $\beta$, recovery rate $\ga$, vaccination rate $\sigma$, immunity waning rate $\alpha$ and balanced birth and death rates $\delta$.}
	\label{fig_SIR-Flow1}
\end{figure}

\bsn
The simplest model to study this question is the so-called SIRS model furnished with an immunity waning rate $\al$ and a vaccination rate $\si$. The model is based on the classic SIR model of \autocite{KerMcKen}, where a population of size $N$ is assumed to be divided into three compartments  $S$ (susceptible), $I$ (infectious) and $R$ (recovered and/or immune) such that $N=S+I+R$. The dynamics of the disease is modeled by an
infection flow from $S$ to $I$, a recovery flow from $I$ to $R$, a loss of immunity flow from $R$ to $S$ and a vaccination flow from $S$ to $R$, see \figref{fig_SIR-Flow1}. 
Simplifying assumption are 
\begin{itemize}
\item[-]
All three compartments are homogeneously mixed within population. 
\item[-]
The average number 
$\beta$ of effective contacts per day (i.e. contacts leading to an infection given the contacted was susceptible) of an infectious person is constant in time and independent of $N$.\footnote{This is the standard incidence assumption. In models with time varying population size $N$ one might also assume a so-called mass-incidence, where $\beta$ is proportional to $N$.} 
So the transmission rate as the (time dependent) number of secondary infections per day caused by a single infectious individual is given by $\beta S/N$.
\item[-]
The incubation time is neglected, i.e. exposed people are considered susceptible.
\item[-]
The time of infectiousness\footnote{Loosely speaking also ``recovery time'', although this is not quite the same.} is distributed exponentially with  mean  time $\Tinf=\ga\inv$, where 
$\ga>0$ is the recovery (more precisely: infectiousness waning) rate.
\item[-]
Recovered persons start immune in $R$, but loss of immunity brings them back to $S$. The duration of immunity is also distributed exponentially with mean duration $\Timm=\al\inv$, where 
$\al>0$ is the immunity waning rate. 
\item[-]
A constant fraction $\si$ of susceptibles gets vaccinated per day. Vaccinated and recovered people behave the same way.
\item[-]
The population size $N$ is assumed constant, so at the end births and deaths are neglected. But to start the discussion more generally, at first I will also include a balanced  demographic birth and death rate $\delta$, where for simplicity the death rate is assumed independent of the compartments  and newborns are assumed susceptible.
\end{itemize}

Given these assumptions one is lead to the following ODE:
\begin{align}
dS/dt &= 
-\be SI/N -(\si+\delta)S + \al R +\delta N\,,
\label{SIRS1}\\
dI/dt &= \be SI/N -(\ga+\delta) I\,, \label{SIRS2}  \\
dR/dt &= \si S+\ga I-(\al+\delta)R\,.  \label{SIRS3}
\end{align}
Note that by construction the total population size is a constant of motion, $dN/dt=0$. In principle one could also consider SIS-type models, where recovery does not lead to immunity and hence flows from $I\rto S$. The methods of this paper would apply to such a model as well \autocite{Nill_SIRS}. More complex models  may also contain an exposed compartment $E$ (SEIRS model) to consider incubation time or a separate vaccination compartment $V$ to distinguish  immunity after vaccination from immunity after recovery. 

\bsn
As has first been shown by \autocite{Hethcote1974} (see also
\autocite{Hethcote1976, Hethcote1989}), for $\delta>0$ already the model without vaccination and loss of immunity, $\al=\si=0$, shows a bifurcation from a stable disease-free equilibrium point (EP) to a stable endemic scenario when raising the basic reproduction number above one. 
The same holds true for $\al>0$ and $\delta\geq 0$, which may be understood intuitively since loss of immunity acts like dying away from $R$ and being newborn into $S$. Nowadays the case 
$\al=\si=0$ and $\delta>0$ is is considered as Hethcote's 
{\em classic endemic model}.

\bsn
Usually endemic models are used for studying diseases over longer periods, during which there is a renewal of susceptibles by births or falling back from temporary immunity causing diseases to return in (damped) periodic waves
\autocite{Hethcote2000}. As contact rates become very large and/or duration of immunity relatively small, the frequency of these waves increases.
Surveys of more general models with periodic behavior are given in \cite{HethStechDries} and \cite{HethLevin}. These models also include nonlinear incidence, temporary immunity by time delay methods or explicitly periodic parameters. Models with varying population size, in particular also disease induced mortality, have been analyzed e.g. in \autocite{BusDries90} and \autocite{Mena-LorcaHeth}. 
When considering vaccination the simplest way is of course statically, i.e. in the sense of initial conditions on the size of the immune compartment $R$ in order to acquire herd immunity. But the shorter the expected duration of immunity the more important becomes the role of constant vaccination models.

\bsn
Meanwhile there are plenty of papers generalizing Hethcote's original ideas, partly also not being aware of each other. The following list is without claim for completeness and with apologies for the unavoidably overlooked ones. 

For SIRS/SIS models without vaccination see e.g. 
\autocite{KorobWake} or \autocite{ORegan_et_al}.
A SIS-version of \eqref{SIRS1}-\eqref{SIRS3} with  varying population size has been analyzed by 
\autocite{LiMa2002} and in \autocite{LiMa2004} the authors have proposed a generalization modeling waning immunity by a time-delay differential equation. In  \autocite{Chauhan_et_al} and \autocite{Batistela_et_al} the authors have added a vaccination term to the classic endemic model, unfortunately without referring to Hethcote's original work. 

Generalizations to SEIR-type models without vaccination have been given e.g. by
\autocite{Li_et_al, LiWang, Korobeinikov2004, LiJin, Korobeinikov2009}. For further generalizations with non-bilinear transmissions see also   
\autocite{LiMuld, KorobMaini, Korobeinikov2006, SunLin}. SEIR-models including vaccination have been analyzed by 
\autocite{SunHsieh} and \autocite{WangXu}.

A model for booster vaccination with a separate compartment for primary vaccination has been proposed by \autocite{Alexander_et_al} and 
periodic pulse vaccination has been studied e.g. in \autocite{LuChiChen, Gao_et_al, ShiDong}. 
Time dependent vaccination programs have also been studied in \autocite{LedzSch} by applying optimal control methods and in \autocite{Kopfova_et_al} by letting the vaccination activity be functionally dependent on the prevalence $I/N$ via the Preisach hysteresis operator.

A different approach to modeling partial and/or waning immunity has been proposed by \autocite{Had_Cast}, where in a combined SIS/SIRS core group model the authors have introduced a diminished transmission rate directly from $R$ to $I$.  
Models with infection transmissions from several compartments may show a so-called {\em backward bifurcation} from the disease-free to an endemic scenario \autocite{Had_Dries}. This means that two (or more) equilibrium states may coexist locally stable for some range below threshold, causing also hysteresis effects upon varying parameters. In 
\autocite{KribsVel} the authors have extended these results to  a combined SIS/SIRS model with vaccination and two immunity waning flows, $R\rto S$ and 
$R\rto I$. Distinguishing vaccinated and recovered people into separate compartments, similar results have been obtained by  \autocite{Arino_et_al}. More recently these ideas have been generalized to a thorough stability analysis of an eight parameter SIRS-type model including 
varying population size in \autocite{AvramAdenane2022, AvramAdenane_et_al}.

Backward bifurcation has lately also been observed in SEIRS-type models for Covid-19 by considering two distinguished susceptible compartments. In \autocite{NadimChatto} the less susceptible compartment had been interpreted as an incomplete lockdown and in \autocite{Diagne_et_al} as an incomplete vaccination efficacy. 
A problem for such models of course arises when trying to decide from empirical reinfection data to which loss/absence-of-immunity model the data should fit (i.e. with a flow $I\rto S\rto I$ (SIS-model) or $I\rto R\rto S\rto I$ (SIRS-model)  or 
$I\rto R\rto I$ (the above models)).

Closing this overview I should also remark that backward bifurcation is also observed when considering $I$-dependent contact or recovery rates to model reactive behavior or infection treatment. However the list of papers on this topic over the last 20 years becomes too huge to be quoted at this place. 

\bsn
In most of the above papers focus is put on questions of stability and thresholds. Although already in Hethcote's original work \autocite{Hethcote1974, Hethcote1976, Hethcote1989} the appearance of a spiraling endemic equilibrium node had explicitly been stated, thresholds  separating the non-oscillating from the damped-oscillating scenario are rarely given explicitly.
Only recently \autocite{Greer_et_al} used a variant of the classic endemic model (i.e. without vaccination, with mass incidence and with unbalanced birth and death rates) to apply such thresholds when analyzing historical smallpox waves. 
For a numerical analysis of a 7-compartment SEIRS-type model with vaccination and waning immunity describing periodic large outbreaks of Mumps in Scotland see
\autocite{Hamami_et_al}.

\bsn
In this paper I will give explicit formulas for the bounds leading to a {\em spiral endemic equilibrium} in
the SIRS model \eqref{SIRS1}-\eqref{SIRS3}. On the way I will also show that this model  in fact maps to Hethcote's classic endemic model by a shift-and-rescaling transformation of variables. More generally such a map also exists for models like e.g. a mixed SIRS/SIS model, models with vaccination rate proportional to $I$
and models with unbalanced birth and death rates, vertical transmission and part of the newborns vaccinated \autocite{Nill_SIRS}.

Based on latest data of the SARS-CoV-2 Omicron  variant I will then argue that according to this simplified model logistically feasible vaccination programs will most likely not be capable to get this epidemic out of an (albeit strongly damped) oscillating endemic phase. Numerical tables in the Appendix support this picture, while at the same time giving evidence that the damping factor most likely will be too strong for these oscillations to be observed empirically.

\section{The classic endemic model \label{Sec_Vacc} }
In this section I will show that for $\beta>0$, $\gd:=\ga+\delta>0$  and all other parameters non-negative the SIRS model \eqref{SIRS1}-\eqref{SIRS3} after a variable transformation looks like the classic endemic model with suitable choices of parameters. 
So as usual, in a first step we measure time in units of $\gd$ and introduce rescaled variables 
\begin{equation}
\tau:=\gd t\,,\qquad r_0:=\beta/\gd\,,
\qquad x:=\frac{r_0 S}{N}\,, \qquad
y:=\frac{r_0 I}{N}\,. \label{xy} 
\end{equation}
Denoting derivatives w.r.t. $\tau$ by dots and replacing $R=N-S-I$ we end up with the two-dimensional system 
\begin{align}
\dot{x}&=-xy- ay-bx+r_0c,
\label{Sir_gen1} \\
\dot{y}&=xy-y,
\label{Sir_gen2}
\end{align}
where the new dimensionless parameters $a, b, c$ are given by
\begin{equation}
a :=\al/\gd\,,
\qquad
b :=(\al+\delta+\si)/\gd\,,
\qquad
c := (\al+\delta)/\gd\,. \label{abc} 
\end{equation}
Before proceeding let me shortly recall the meaning of $r_0$ and 
$x$. First, according to the standard definition (see e.g. \autocite{Hethcote2000} or \autocite{Anderson_May}) in models containing just one infectious compartment the {\em basic reproduction number} $r_0$ is given as the expected number of secondary cases produced by a typical infectious individual in a completely susceptible population $S=N$. So this is the effective contact rate $\beta$ times the mean time of infectiousness and therefore, in the presence of a death rate, 
$r_0=\beta/(\ga+\delta)$, in consistency with \eqref{xy}. 

Second, according to \autocite{Hethcote2000} the {\em replacement number} $x$ as a function of time is defined to be the expected number of secondary cases produced by a typical infectious individual during its time of infectiousness. Hence  $x$ is given by $r_0$ times the probability of a contact being susceptible\footnote{Strictly speaking one should average this probability over the time of infectiousness, but on this time scale $S/N$ may safely be assumed constant.}, $x=r_0S/N$, which coincides with the definition in \eqref{xy}. Nowadays the replacement number is mostly called {\em effective reproduction number}, but this might lead to misunderstandings, since there is also a notion of a {\em vaccination-reduced reproduction number $\R_0$} as a threshold  parameter to be explained in Appendix \ref{Sec_threshold}. 

\bsn
Coming back to the parameters in \eqref{abc}, note that they satisfy the constraints
\begin{align}
0 &\leq c\leq  b\,,\label{constraint_1} \\
a &\leq c \leq 1+ a\,.\label{constraint_2}
\end{align} 
If one didn't look at \eqref{abc} then from \eqref{constraint_1} and \eqref{constraint_2} one would also conclude
\begin{equation}
-1 \leq a\leq b\,.\label{constraint_3}
\end{equation}
Now by definition $a$ seems to be non-negative. But in fact, assume in place of the SIRS model \eqref{SIRS1}-\eqref{SIRS3} we had started with the analogous SIS model. Then we would also end up with the system \eqref{Sir_gen1}-\eqref{Sir_gen2}, but in this case the definition of $a$ would be replaced by
\begin{equation}
a:=(\al-\ga)/\gd=c-1\geq -1\,.
\end{equation}
So in this way we may consider the system \eqref{Sir_gen1}-\eqref{Sir_gen2} for $(x,y)\in\RR_{\geq 0}^2$ and with constraints \eqref{constraint_1}-\eqref{constraint_3} as a master system covering all models of type SIRS or SIS (or mixed) as in \eqref{SIRS1}-\eqref{SIRS3}, with vaccination rate $\si\geq 0$ and immunity waning rate $\al\geq 0$. In particular the classic endemic model corresponds to $a=0$ and 
$0<b=c=\delta/(\ga+\delta)<1$.

\bsn
Moreover, it is not difficult to check, that the {\em physical triangle} given by $S+I+R=N$ or equivalently 
\begin{equation} 
\Tph=\{(x,y)\in\RR_{\geq 0}^2\mid x+y\leq r_0\}
\label{phys_states} 
\end{equation}
stays forward invariant under the dynamics \eqref{Sir_gen1}-\eqref{Sir_gen2} provided the constraints 
\eqref{constraint_1} - \eqref{constraint_3} hold.

\bsn
In the second step I am now going to show that except for the border case 
$a=-1$\footnote{This corresponds to a SIS model with $\si\geq 0$ and $\al=\delta=0$, which epidemiologically is uninteresting.} we may  always rescale to $a=0$. In fact, 
there still is a combined ``space-time'' scaling redundancy in the system \eqref{Sir_gen1}-\eqref{Sir_gen2} given by the one-parameter group of variable transformations
$$(x-1)\mapsto\lambda(x-1)\,,\qquad
y\mapsto\lambda y\,,\qquad
\tau\mapsto\lambda\inv\tau\,,
\qquad\lambda>0\,.
$$
This leaves the system \eqref{Sir_gen1}-\eqref{Sir_gen2} invariant provided the parameters $a,\,b,\,r_0c$ are also rescaled according to
$$
(a+1)\mapsto \lambda(a+1)\,,\qquad
b\mapsto\lambda b\,,\qquad (r_0 c-b)\mapsto\lambda^2(r_0 c-b)\,.
$$
So for $a>-1$ this leads to introducing adapted ``normalized'' variables 
\begin{equation}
u(\tilde{\tau}):=\frac{x(\tau)+a}{1+a}\,,\quad
v(\tilde{\tau}):=\frac{y(\tau)}{1+a}\,,\quad
\tilde{\tau}:=(1+a)\tau\,.\label{uv-variables} 
\end{equation}
In terms of these variables the equations of motion become
\begin{align}
\dot{u}&=-uv-c_1u+c_2\,,\label{dot_u} \\
\dot{v}&=uv-v\,,\label{dot_v} 
\end{align}
where now dots denote derivatives w.r.t. $\tilde{\tau}$ and where the new parameters are given by
\begin{align}
c_1&=b/(1+a)\geq 0\,,
\label{c_1} \\
c_2&=(ab+r_0c)/(1+a)^2=c_1+(r_0c-b)/(1+a)^2\in\RR\,.\label{c_2}
\end{align}
Apparently for $c_1=\delta/(\ga+\delta)$ and $c_2=r_0 c_1$ we  precisely recover the classical endemic model. The price to pay is that in the SIS-model-case we may have $a<0$ and hence possibly also negative values of $u$ and $c_2$. Thus, in order to cover the most general setting we have to consider \eqref{dot_u}-\eqref{dot_v} as a dynamical system on phase space 
$(u,v)\in\RR\times\RR_{\geq 0}$ and the admissible range of parameters is 
$(c_1,c_2)\in (\RR_+\times\RR)\,\cup\, \{(0,0)\}$. In fact, under these conditions the master system \eqref{dot_u}-\eqref{dot_v} also covers more general models like e.g. a mixed SIRS/SIS model, models with vaccination rate proportional to $I$, and models with unbalanced birth and death rates, vertical transmission and part of the newborns vaccinated \autocite{Nill_SIRS}.

\section{The main Theorem\label{Sec_main-theorem} }
Having reduced a whole class of models to a (marginally extended) version of Hethcote's classic endemic model standard results  now easily carry over. First note that  the case $c_1=0$ means $\si=\al=\delta=c_2=0$ and hence reduces to the classical SIR or SIS model, which here I am not interested in. So from now on assume $c_1>0$ or equivalently $\al+\delta+\si>0$. 

\bsn
Now it is important to realize, that given $c_1>0$ and $c_2\in\RR$ any initial value
$(u_0,v_0)\in\RR\times\RR_{\geq 0}$ for the dynamical system \eqref{dot_u}-\eqref{dot_v} may be considered to lie in the image of some physical triangle $\Tph$ under the transformation \eqref{uv-variables} and \eqref{c_1}-\eqref{c_2}. 
Thus for any initial value $(u_0,v_0)$ the forward time evolution 
$(u(\tilde{\tau}),v(\tilde{\tau}))$ under the dynamics \eqref{dot_u}-\eqref{dot_v} stays bounded and exists for all 
$\tilde{\tau}>0$ \autocite{Nill_SIRS}. This allows to apply standard techniques by using Lyapunov functions and LaSalle's Invariance Principle, see e.g. \autocite{Hethcote1989} or \autocite{Mena-LorcaHeth}.

\bsn
From now on the way to proceed is straight forward. Writing the master system \eqref{dot_u}, \eqref{dot_v} in the form 
$\dot{\bp}=\X(\bp)$ equilibrium points $\bp^*$ are given as zeros of the vector field, $\X(\bp^*)=0$. There are precisely two solutions 
$\bp_i^*=(u_i^*,v_i^*)$, $i=1,\,2$, given by 
\begin{align}
u^*_1&=\frac{c_2}{c_1}, \qquad v^*_1=0,\label{uv_1*} \\
u^*_2&=1, \qquad\ \, v^*_2=c_2-c_1.\label{uv_2*} 
\end{align}
In coordinates $(x,y)$ they correspond to 
\begin{align}
x^*_1&=\frac{r_0c}{b}, \qquad y^*_1=0,\label{xy_1*} \\
x^*_2&=1, \qquad\quad  y^*_2=\frac{r_0c-b}{1+a}\,,\label{xy_2*} 
\end{align}
and in terms of the original SIRS-model variables for $\delta=0$
\begin{align}
S_1^*/N&=\frac{\al}{\al+\si}\,,\qquad I_1^*=0\,,
\qquad\qquad\qquad\qquad\ R_1^*/N=\frac{\si}{\al+\si}\\
r_0S_2^*/N &=1\,,\qquad 
r_0I_2^*/N =\frac{(r_0-1)\al-\si}{\ga+\al}\,,\qquad
r_0R_2^*/N=\frac{(r_0-1)\ga+\si}{\ga+\al}\,.
%
\label{SI_2*} 
\end{align}
For $c_2=c_1$ the two EPs coincide, $\bp_1^*=\bp_2^*$. As we will see, this threshold marks the transition from the stable disease-free to the stable endemic equilibrium. This motivates to distinguish the following three scenarios (A) - (C)

\begin{equation}
\begin{array}{rccccccccl}
(A):\quad &\vs < 0 &\LRA& \us < 1 	&\LRA& c_2 < c_1
	&\LRA &x_1^*\equiv r_0c/b < 1	&\LRA& r_0<1+\si/\al\,,\\
(B):\quad &\vs = 0 &\LRA& \us = 1   &\LRA& c_2 = c_1
	&\LRA &x_1^*\equiv r_0c/b = 1	&\LRA& r_0=1+\si/\al\,,\\
(C):\quad &\vs > 0 &\LRA& \us > 1   &\LRA& c_2 > c_1
	&\LRA &x_1^*\equiv r_0c/b > 1	&\LRA& r_0>1+\si/\al\,.\\
\end{array}
\label{scenarios} 
\end{equation}
Here for simplicity the last equivalences are expressed for the case $\delta=0$. Next, local asymptotic behavior near the EP 
$\bp_i^*$ is determined by the eigenvalues of the linearized system at $\bp_i^*$. 
Denoting $T_i$ the trace and $D_i$ the  determinant of the Jacobian $D\X(\bp_i^*)$ and putting $\Del_i:=T_i^2-4D_i$  we get
\begin{align}
T_1&=c_2/c_1-1-c_1\,,&
T_2&=-c_2\,,\\
D_1&=c_1-c_2\,,&
D_2&=c_2-c_1\,,\\
\Del_1&=(c_2/c_1 -1+c_1)^2\,,&
\Del_2&=c_2^2-4c_2+4c_1\,.
\label{Del_12} 
\end{align}
Thus the above scenarios (A) and (C) subdivide into
\begin{equation}\label{sub_scenarios} 
\begin{array}{rll}
(A1):& c_2\neq c_1-c_1^2 \land c_2<c_1
& \Longleftrightarrow\ (A)\land \Del_1>0\,,
\\
(A2):& c_2= c_1-c_1^2 \neq 0
&\mspace{-16mu}
\rdelim\}{2}{*}[$\,\Longleftrightarrow\ (A) \land \Del_1=0\,, $]
\\
(A3):& c_2=0\land c_1=1
& 
\\
(C1): & c_2-c_2^2/4<c_1<c_2
& \Longleftrightarrow\ (C)\land \Del_2>0\,,
\\
(C2): & c_2-c_2^2/4=c_1
& \Longleftrightarrow\ (C)\land  \Del_2=0\,,
\\
(C3): & c_2-c_2^2/4>c_1
& \Longleftrightarrow\ (C)\land  \Del_2<0\,.
\end{array}
\end{equation}
The following unifies various results in the literature as quoted in the introduction.
{\theorem\label{Thm_Stability} 
For $(c_1,c_2)\in\RR_+\times\RR$ consider the master system \eqref{dot_u}-\eqref{dot_v} on $\RR^2$.
\begin{itemize}
\item[i)]
In scenario $(A)$ the EP $\bp_2^*=(1,c_2-c_1)$ is an (unphysical) saddle point and the disease free EP 
$\bp_1^*=(c_2/c_1,0)$ is a stable node which is proper in 
$(A1)$, degenerate in $(A2)$ and star in $(A3)$. 
\item[ii)]
In scenario $(B)$ the two equilibria coincide, 
$\bp_1^*=\bp_2^*=(1,0)$ and this EP is non-hyperbolic.
\item[iii)]
In scenario $(C)$ the disease free EP $\bp_1^*$  is a saddle point and the endemic EP $\bp_2^*$ is a stable node which is proper in 
$(C1)$, degenerate in $(C2)$ and spiral in $(C3)$.
\item[iv)]
In scenarios $(A)$ and $(B)$ the closed upper half-plane 
$\{v\geq 0\}$ is an asymptotic stability region for 
$\bp_1^*$ and in scenario $(C)$ the open upper half-plane 
$\{v>0\}$ is an asymptotic stability region for 
$\bp_2^*$.
\end{itemize}
}
\proof
Parts i)-iii) immediately follow from the definitions \eqref{sub_scenarios} and the eigenvalue formulas
\begin{equation}\label{eigenvalue} 
\lambda_{i,1/2}=\frac{1}{2}\left(T_i\pm\sqrt{\Del_i}\right),
\qquad T_i=\lambda_{i,1}+\lambda_{i,2},\qquad 
D_i=\lambda_{i,1}\lambda_{i,2}\,.
\end{equation}
To prove part iv) one may adapt standard arguments using Lyapunov functions and LaSalle's Invariance Principle, see e.g. \autocite{Hethcote1989} or \autocite{Mena-LorcaHeth}.
A complete proof will be given in \autocite{Nill_SIRS}.
\qed

\bsn
Computing eigenvectors also yields asymptotic slopes 
$\left(\dot{v}/\dot{u}\right)_\infty$ at the EPs. A complete overview is given in
Table \ref{tab:1}. 
Here in the case of proper nodes orbits are called ``generic'' if they  are  asymptotically tangent to the leading eigenvector. So these are almost all orbits except exactly two  tangent to the subleading eigenvector.\footnote{For example, if in scenario 
(A1) the leading eigenvalue is given by $\lambda_{1,2}\equiv c_2/c_1-1>\lambda_{1,1}\equiv -c_1$ then all orbits with initial condition $v_0>0$ will obey 
$\left(\dot{v}/\dot{u}\right)_\infty=
(c_1-c_1^2-c_2)/c_2\neq 0$, whereas an initial condition $v_0=0$ will yield $v_\tau=0$ for all $\tau\in\RR$.} 

\begin{table}[htpb!]
\caption{Stable Equilibrium Points (EP)}
\label{tab:1} 
\renewcommand{\arraystretch}{1.4}
\begin{tabular}{|l|l|c|c|c|c|}
\hline
\multicolumn{2}{|c|}{Scenario/Type} & EP & Eigenvalues $\lambda_{i,1/2}$ & 
Asympt. Slope & Conditions
\\
\hline\hline
\multirow{2}{*}{A1} &
\multirow{2}{1cm}{proper} & 
\multirow{6}{*}
{\rotatebox{90}{$(u_1^*,v_1^*)=(\frac{c_2}{c_1},0)$}} &
$\lambda_{1,1}=-c_1$ & $0$ & \rule[-3mm]{0mm}{11mm}
$
\begin{gathered}
c_2<c_1-c_1^2 \\
\text{(generic orbit)}
\end{gathered}
$
\\
\cline{4-6}
& & & $\lambda_{1,2}=c_2/c_1 -1$ & \rule[-3mm]{0mm}{11mm}
$\displaystyle \frac{(c_1-c_1^2-c_2)}{c_2}$ & 
$
\begin{gathered}
c_1>c_2>c_1-c_1^2 \\
(v_0>0)
\end{gathered}
$
\\
\cline{1-2}\cline{4-6}
A2 & degenerate & &
$\lambda_{1,1/2}=-c_1\neq-1$&$0$&$c_2=c_1-c_1^2\neq 0$
\\
\cline{1-2}\cline{4-6}
A3 & star & & $\lambda_{1,1/2}=-1$ & any value & $c_1=1,\ c_2=0$
\\
\cline{1-2}\cline{4-6}
\multirow{2}{*}{B} & \multirow{2}{2cm}{non-hyperbolic} & &
$\lambda_{1,1}=-c_1$ & $0$ & 
$
\begin{gathered}
c_1=c_2\ 
(v_0=0)
\end{gathered}
$
\\
\cline{4-6}
& & & $\lambda_{1,2}=0$ & $-c_1$ & 
$
\begin{gathered}
c_1=c_2\ 
(v_0>0)
\end{gathered}
$
\\
\hline
\multirow{2}{*}{C1} & 
\multirow{2}{1.8cm}{proper} & 
\multirow{4}{*}[2.4mm]
{\rotatebox{90}{$(u_2^*,v_2^*)=(1,c_2-c_1)$}} &
$\lambda_{2,1}=\frac{1}{2}(-c_2+\sqrt{\Del_2})$ 
& $-\frac{1}{2}(c_2+\sqrt{\Del_2})$  
& \rule[-3mm]{0mm}{11mm}
$
\begin{gathered}
0<\Del_2<c_2^2 \\
\text{(generic orbit)}
\end{gathered}
$
\\
\cline{4-6}
& & & $\lambda_{2,2}=\frac{1}{2}(-c_2-\sqrt{\Del_2})$ & 
$-\frac{1}{2}(c_2-\sqrt{\Del_2})$  
& \rule[-3mm]{0mm}{11mm}
$
\begin{gathered}
0<\Del_2<c_2^2 \\
\text{(special orbit)}
\end{gathered}
$
\\
\cline{1-2}\cline{4-6}
C2 & degenerate & &
$\lambda_{2,1/2}=-c_2/2$ & $-c_2/2$ &$\Del_2=0$
\\
\cline{1-2}\cline{4-6}
C3 & spiral & & $\lambda_{2,1/2}=\frac{1}{2}(-c_2\pm\sqrt{\Del_2})$ & none & $\Del_2<0$
\\
\hline
\end{tabular}
\renewcommand{\arraystretch}{1}
\end{table}

\section{The oscillating endemic scenario}
By \Eqref{scenarios} the threshold for endemic bifurcation is given by
\begin{equation}
r_0>b/c = 1+\si/\al\,,\label{r_end} 
\end{equation}
where the second equality holds for $\delta=0$. So in this section I will focus on the thresholds for the oscillating endemic scenario (C3).
First note that the condition for spiraling, 
$c_2-c_2^2/4>c_1$, necessarily requires $c_1<1$ or equivalently 
$b<1+a$. Sufficiency is obtained by requiring also lower and upper bounds on $r_0$. Put
\begin{equation}
r_\pm :=\frac{b}{c}+\frac{1+a}{c}
\left(\sqrt{1+a}\pm\sqrt{1+a-b}\right)^2\,.
\label{rpm1} 
\end{equation}
{\corollary{\label{Cor_C3}}
Scenario (C3) is equivalent to $b<1+a$ and $r_-<r_0<r_+$.
}
\proof
Using Eqs. \eqref{c_1}, \eqref{c_2} and \eqref{Del_12} we have
\begin{equation}
\Del_2=c_2^2-4c_2+4c_1=\frac{c^2}{(1+a)^4}(r_0-r_-)(r_0-r_+)\,.
\end{equation}
\qed

Asymptotic values for the decay half-life $\Thalf$ and the oscillation period $\Tosc$ in scenario (C3) can now be read off from the real/imaginary part of the eigenvalues (last line of Table \ref{tab:1}). Recalling
$\tilde{\tau}=(1+a)\ga t$ this gives 
\begin{align}
\ga \Thalf&=\frac{2\log 2}{(1+a)c_2}
= \frac{2\log 2\, (1+a)}{r_0c+ab}\,,
\label{Thalf} 
\\
\ga\Tosc &=\frac{4\pi}{(1+a)\sqrt{-\Del_2}}
=\frac{4\pi(1+a)}{\sqrt{-(r_0c+ab)^2+4(1+a)^2(r_0c-b)}}\,.
\label{Tosc} 
\end{align}
Let us now apply this to the SIRS model without vital dynamics, 
$\delta=0$. As may be seen from the tables in  Appendix \ref{Sec_SIRS-tables} (see \figref{Fig_time_scales}), for a wide range of parameters  $\Tosc$ will roughly be 5 times bigger than 
$\Thalf$. Hence, in the course of one wave cyle amplitudes get already damped by a factor of roughly 0.05. So empirically these waves would most likely be swallowed by noise effects and hence presumably not be observable.

\bsn
Moreover, for $\delta=0$ we have $a=c=\al/\ga$ and $b=a+\rv$ where $\rv:=\si/\ga$. Hence 
\begin{equation}
\delta=0\quad\Longrightarrow\quad
 r_\pm =1+\rv/a + (1+a\inv)(\sqrt{1+a}\pm\sqrt{1-\rv})^2\,.
\label{rpm2} 
\end{equation}
Also, in this case $b<1+a$ is equivalent to $\rv<1$. 
Note that for $\rv\in[0,1]$ we have 
$\pm \partial r_\pm /\partial \rv<0$
and therefore the interval $[r_-,r_+]$ gets narrower as $\rv$ increases. Let me call $\rv$ the {\em vaccination activity}. As will be seen in the next Section for Covid-19 we may safely assume $\rv < 1$.


\section{Numerical estimates}
To get numerical input we now need estimates for 
$\ga, a$ and $\rv$. Since the SIRS model is much too simple to describe reality quantitatively, I will only go for rough estimates to get a feeling for orders of magnitude. The aim is to see,  whether empirical data are far from thresholds so the model's qualitative predictions may be judged realistic.

\bsn
Let us first look at latest studies estimating the mean time of infectiousness, $T_{inf}=\ga\inv$. On 2021-12-22 the UK Health Security Agency (UKHSA) gave  new guidance for the public and health and social care staff. In \autocite{UKHSA_20220110} the agency quotes a recent  modeling study
\autocite{Bays_et_al}, according to which after 10 full days of self-isolation 5\% of people who tested
positive for SARS-CoV-2 are still infectious. Numbers reported are also 15.8\% after  7 days and 31.4\% after 5 days. 
Mapping these data to an exponential decay as assumed by the SIR model one gets $\ga\approx$  0.30 - 0.23 corresponding to 
$T_{inf}\approx$ 3.4 - 4.3 days. The above data do not include the Omicron variant of SARS-CoV-2. Mostly Omicron seems to be less severe then Delta indicating shorter recovery times. But in lack of better knowledge let's stay conservative and assume the same range for Omicron.

\bsn
Concerning estimates on the expected duration of immunity,
$T_{imm}=\al\inv$, actual studies for Omicron are still volatile and ongoing. For almost weekly updates see e.g. the 
UKHSA technical briefing documents
\footnote{\url{www.gov.uk/government/publications/investigation-of-sars-cov-2-variants-technical-briefings}} and the
COVID-19 vaccine weekly surveillance reports\footnote{\url{www.gov.uk/government/publications/covid-19-vaccine-weekly-surveillance-reports}}.
In its technical briefing no. 34 from Jan. 2022 the UKHSA says
{\em ``estimates suggest that vaccine effectiveness against symptomatic disease with the Omicron variant is significantly lower than compared to the Delta variant and wane rapidly''} \autocite{UKHSA_tech-brief34}. 
In a preprint from Dec. 2021 \autocite{Andrews_et_al} state
{\em ``findings indicate that vaccine effectiveness against symptomatic disease with the Omicron variant is significantly lower than with the Delta variant''}. 
Similar findings have also been reported, e.g., by a danish study
\autocite{Lyngse_et_al.} and in Germany by the STIKO recommendation from 2021-12-21 \autocite{Stiko_Vacc}.

Measuring ``effectiveness'' quantitatively numbers of course depend on the specific vaccine. In \autocite{UKHSA_tech-brief34} it is said that among those who had received 2 doses of Pfizer or Moderna effectiveness dropped from around 65 to 70\% down to around 10\% by 20 weeks after the 2nd dose. Two to 4
weeks after a booster dose vaccine effectiveness ranged from around 65 to 75\%, dropping to
55 to 65\% at 5 to 9 weeks and 45 to 50\% from 10+ weeks after the booster. Also, at least against Omicron,  effectiveness apparently never goes above 75\%. Of course the discussion also depends on details like asymptomatic vs. little symptoms or hospitalization etc.

The reinfection risk against Omicron after recovery from Delta also seems to be considerably higher as estimated earlier for Delta - Delta reinfection.
Studies published in 2021 still estimated the anti-SARS-CoV-2-directed IgG-antibody half-life between 85 and 160 days
\autocite{Lumley_et_al, DanMat_et_al, Hartog_et_al}.
But in \autocite{Tounsend_et_al} authors already claimed their results {\em ``indicate that reinfection after natural recovery from COVID-19 will become increasingly common''}. In its report no. 49 from Dec. 2021 the Imperial College Covid-19 response team found {\em ``strong evidence of immune evasion, both from natural infection, where the risk of reinfection is 5.41 (95\% CI: 4.87-6.00) fold higher for Omicron than for Delta, and from vaccine-induced protection} \autocite{ImpCollege_49}, see also
\autocite{Andrews_et_al}.

\bsn
So assuming a simple 1-parameter exponential distribution for immunity as in the SIRS model doesn't quite map the above complexity. Neither does the model distinguish different vaccines nor virus variants nor immunity responses by vaccination vs. recovery. Thus I will plot formulas by assuming a range between 1 and 6 months for the mean duration of immunity $\Timm$, which should be wide enough to cover all reasonable scenarios.
Measuring time in units of $\Tinf$ this gives 
$\Timm/\Tinf\equiv a\inv\approx$ 7 - 53.

\begin{figure}
[htbp]
\centering
	\subcaptionbox{\url{https://coronavirus.data.gov.uk/details/vaccinations}}
	{\includegraphics[width=0.48\textwidth]
	{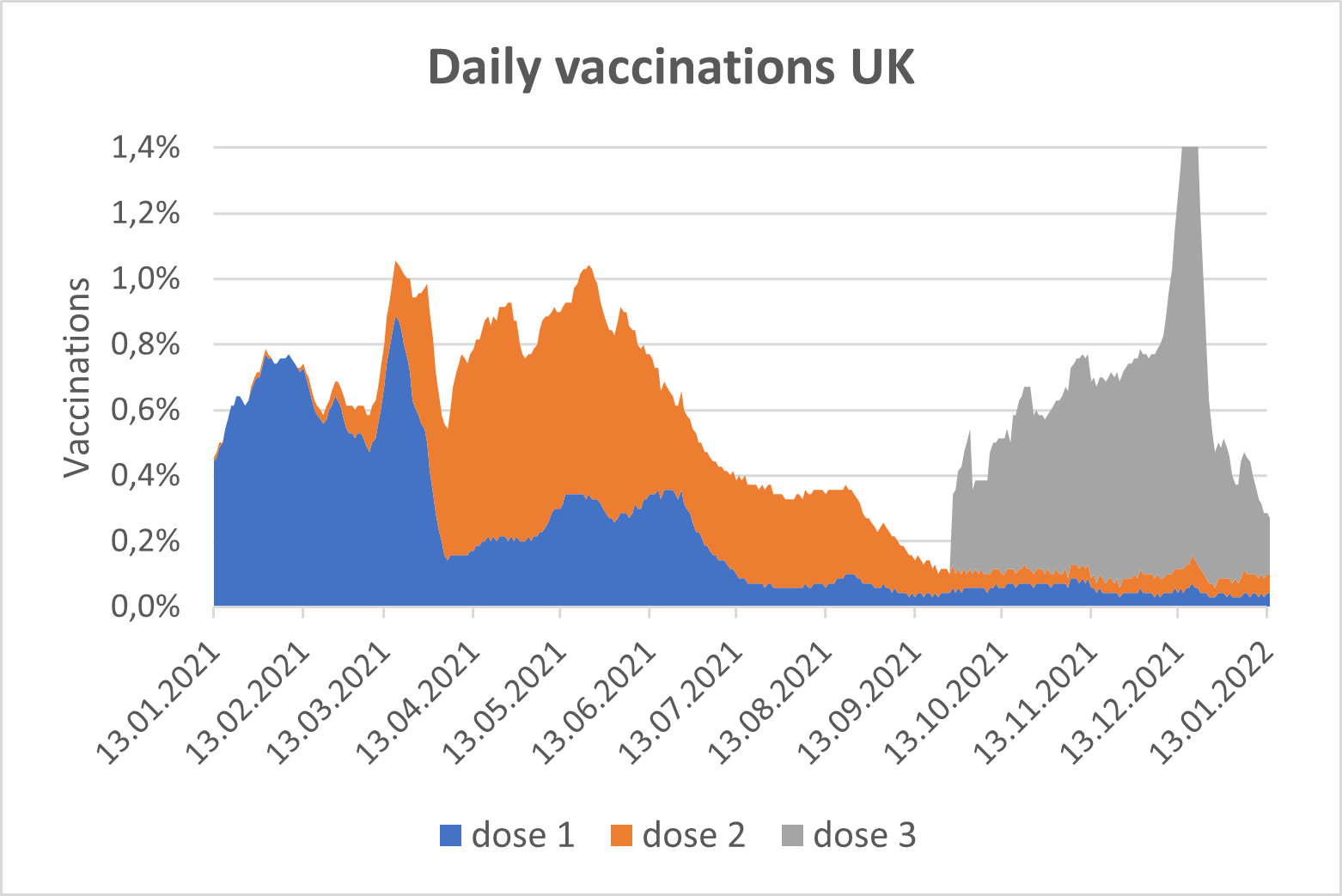}}
	\label{Fig_Vacc_UK}
	%
	\subcaptionbox{\url{https://www.data.gv.at/covid-19/}}
	{\includegraphics[width=0.48\textwidth]
	{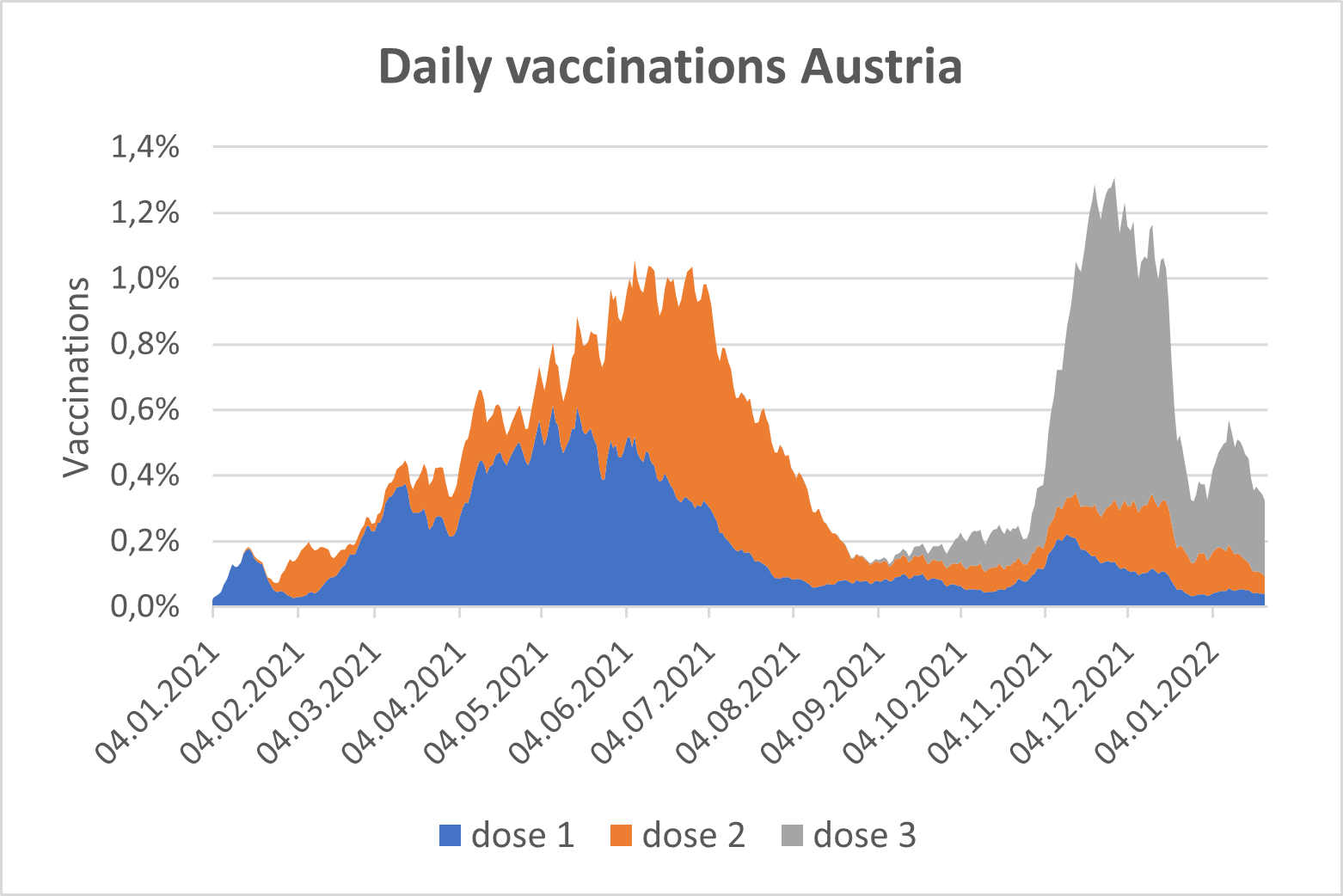}}
	\label{Fig_Vacc_Aus}
	\subcaptionbox{\url{https://www.rki.de/DE/Content/InfAZ/N/Neuartiges_Coronavirus/Daten/Impfquoten-Tab.html}}
	{\includegraphics[width=0.48\textwidth]
	{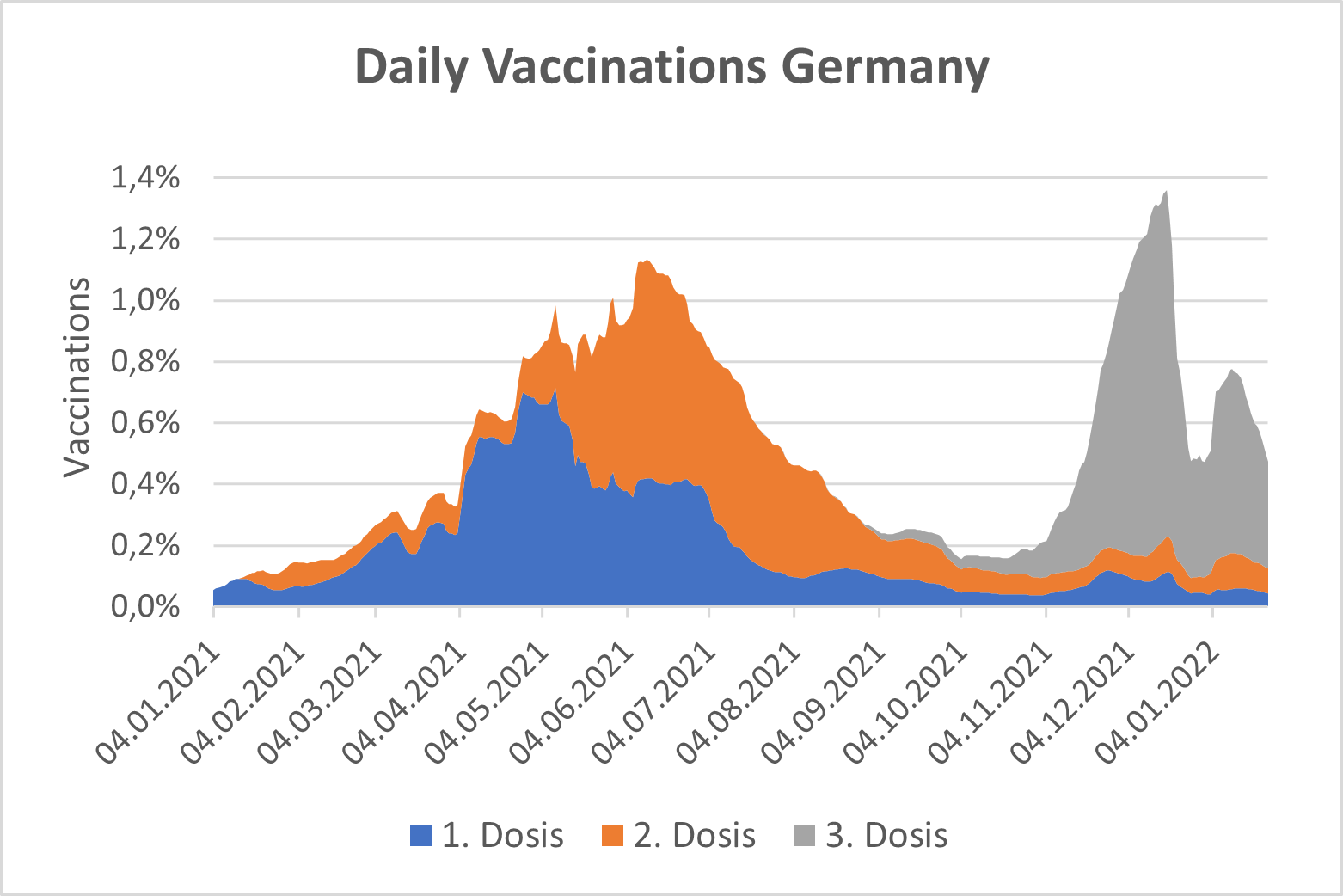}}
	\label{Fig_Vacc_Ger}
	\caption{Daily sum of vaccinations over all dose 1 - 3
	shots in fractions of population.}
\label{Fig_vacc} 
\end{figure}

\bsn
Finally we need an upper bound for the vaccination activity 
$\rv=\si/\ga$. \figref{Fig_vacc} shows public 
data for daily vaccination numbers normalized as fractions of the total population in UK, Germany and Austria.
As a common conclusion the daily sum over all dose 1 - 3 shots rarely ever reaches 1\% of the population. 

Since at that time for most countries a lower bound on the fraction of susceptibles $S/N\gtrsim 0.25$ seems reasonable we get $\si<0.04$ and therefore $\rv<0.17$ as an upper bound which at least as a time-average should safely hold. In particular $\rv<1$ without any doubt, thus assuring $b<1+a$ as the necessary condition for the oscillating scenario (C3), see Corollary \ref{Cor_C3}.

\begin{figure}
[htbp]
\centering
	\subcaptionbox{Lower bounds $r_-$.}[.48\linewidth]		
	{\includegraphics[width=0.48\textwidth]
	{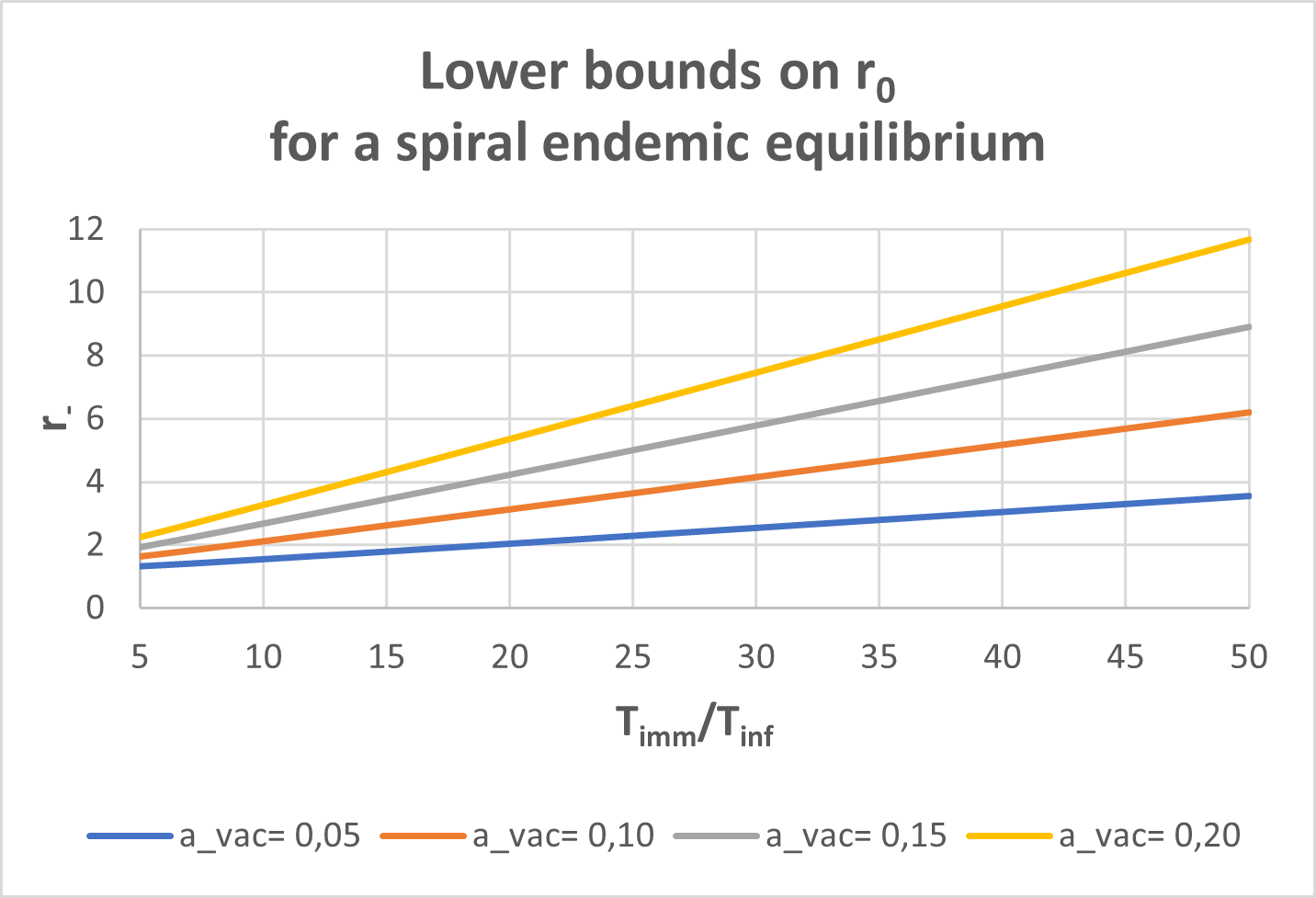}}
	\label{Fig_r-}
	%
	\subcaptionbox{Upper bounds $r_+$}[.48\linewidth]	
	{\includegraphics[width=0.48\textwidth]
	{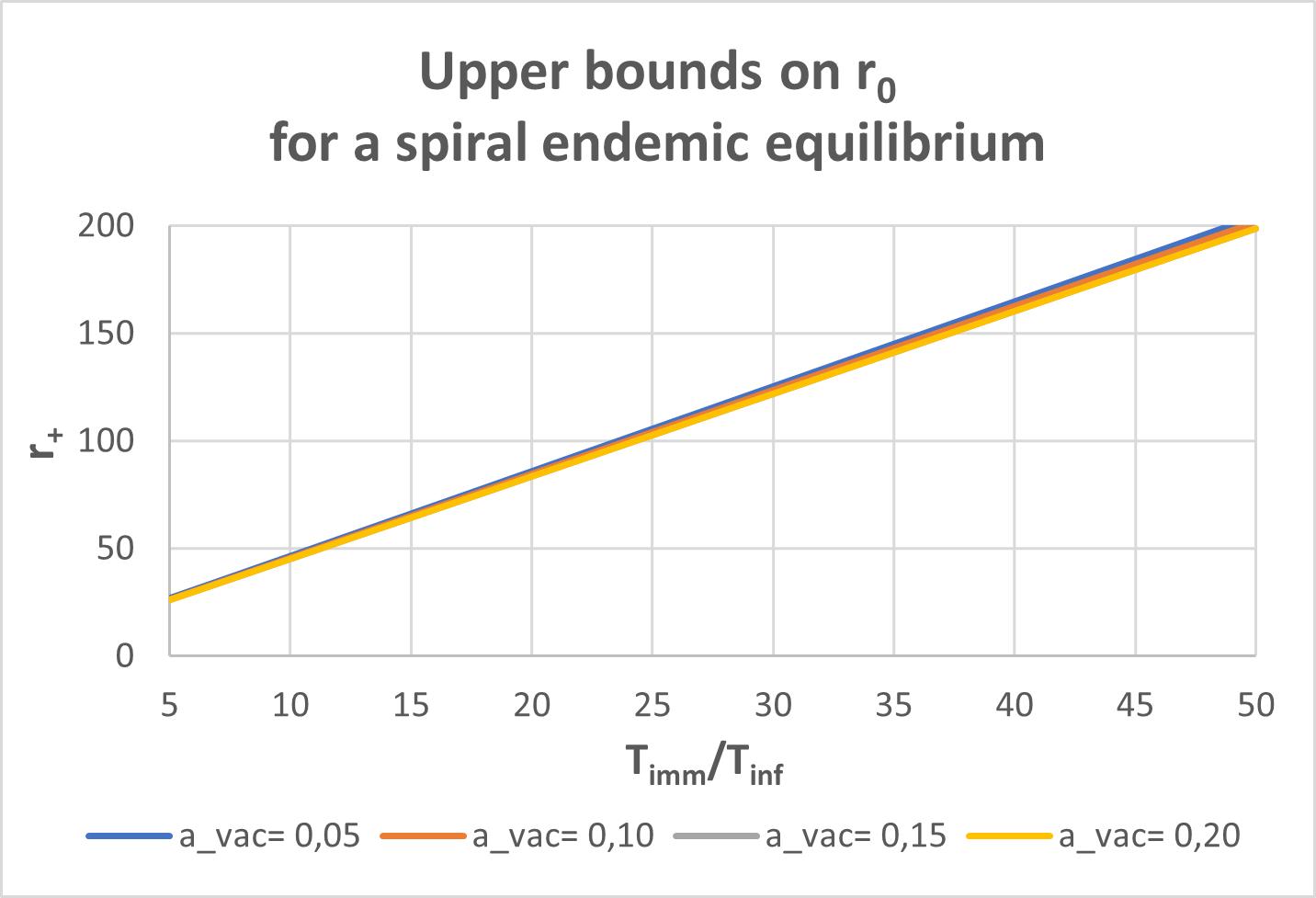}}
	\label{Fig_r+}
	\subcaptionbox{Ratio $r_-$ to endemic threshold $b/c$}[.48\linewidth]	
	{\includegraphics[width=0.48\textwidth]
	{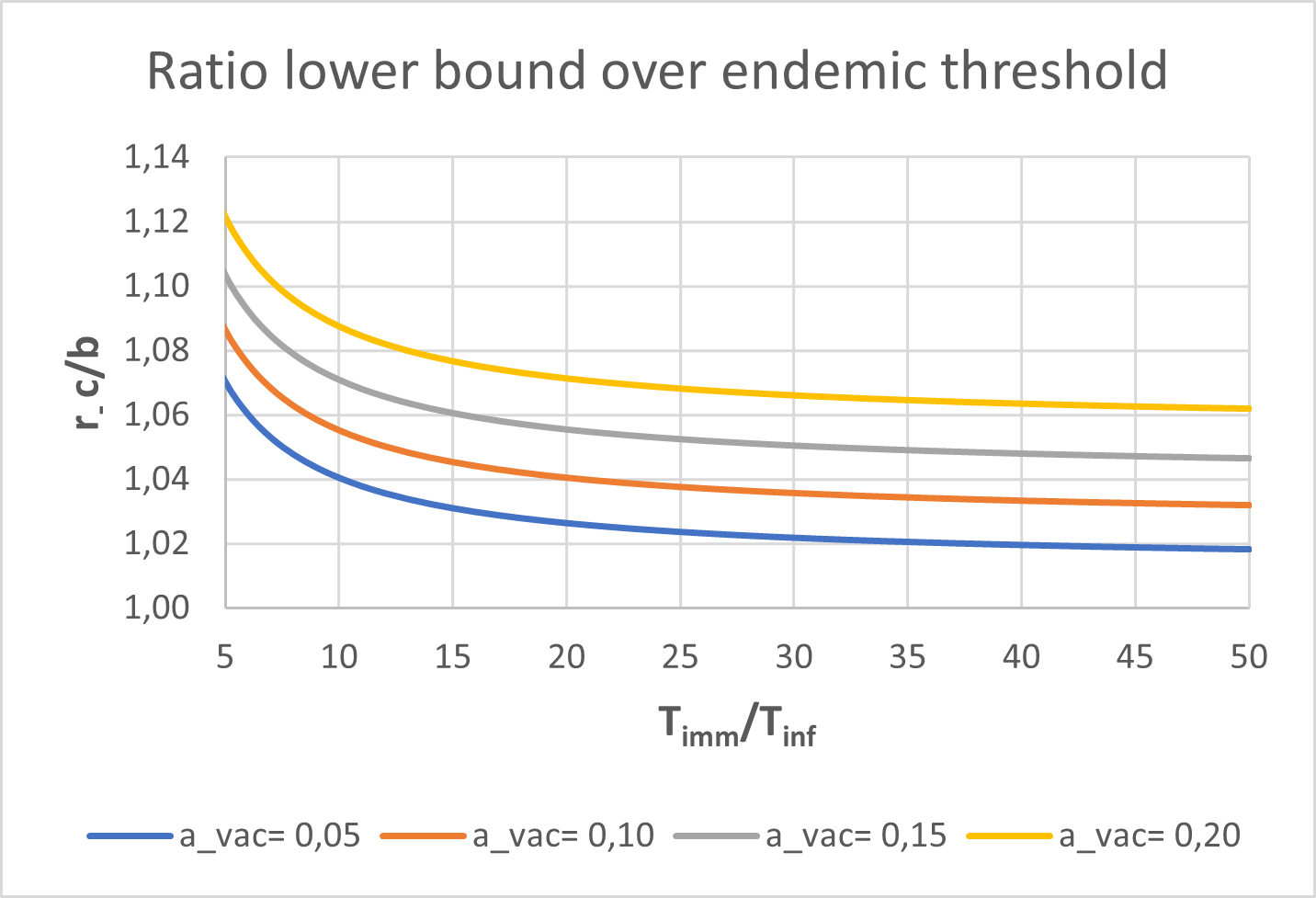}}
	\label{Fig_r-/r_end}
\caption{Lower and upper bounds on $r_0$ for a spiral endemic scenario. Fig. c) shows that for realistic parameter ranges the gap between the endemic threshold $b/c=1+\rv/a$ and the lower bound $r_-$ for spiraling stays below 6\%.}
\label{Fig_r+-} 
\end{figure}

\bsn
In \figref{Fig_r+-} lower and upper bounds 
$r_\pm$ for scenario (C3) 
are plotted over the range $\Timm/\Tinf\in[5,50]$ for values 
$\rv=0.05,\, 0.1,\, 0.15$ and $0.2$. So these lines represent parameter regions for scenario (C2) separating the oscillating endemic scenario (C3) inside $[r_-,r_+]$ from the non-oscillating endemic scenario (C1) outside. 

At first it is obvious, that the upper bound $r_+$ realistically will never be reached. The conclusion from the lower bound is that for $r_0\geq 10$ and a mean duration of immunity $\Timm< 160$ days 
($\Timm/\Tinf< 40$) it seems hardly possible to escape scenario (C3) (leave alone scenario (C)) by manageable vaccination activities. Lowering the assumption on $\Timm$ by 20 days roughly reduces the lower bound on $r_0$ by 1. Also the range between $r_-$ and the threshold $b/c$ marking the border line to scenario (A) is rather narrow. For better visualization a plot of $r_-c/b$ is given in \figref{Fig_r+-}c).

\section{Summary\label{Sec_summary} }

In this paper I have shown that SIRS models (and also SIS models) with constant total population and constant vaccination and immunity waning rates (and possibly also with vital dynamics parameters) may be mapped to Hethcote's classic endemic model, which originally had been based on a balanced birth and death rate only. The only price to pay is an enlarged range of parameter values $c_1, c_2$ and (coming from the SIS model) the possibility of negative values for the would-be replacement number variable $u$. However, these generalizations do not influence the phase structure for equilibrium and stability. Original proofs easily generalize to this master model, thus unifying lots of follow-up proofs on the above models.  

\begin{figure}[htbp!]
\centering
\includegraphics[width=0.98\textwidth]
	{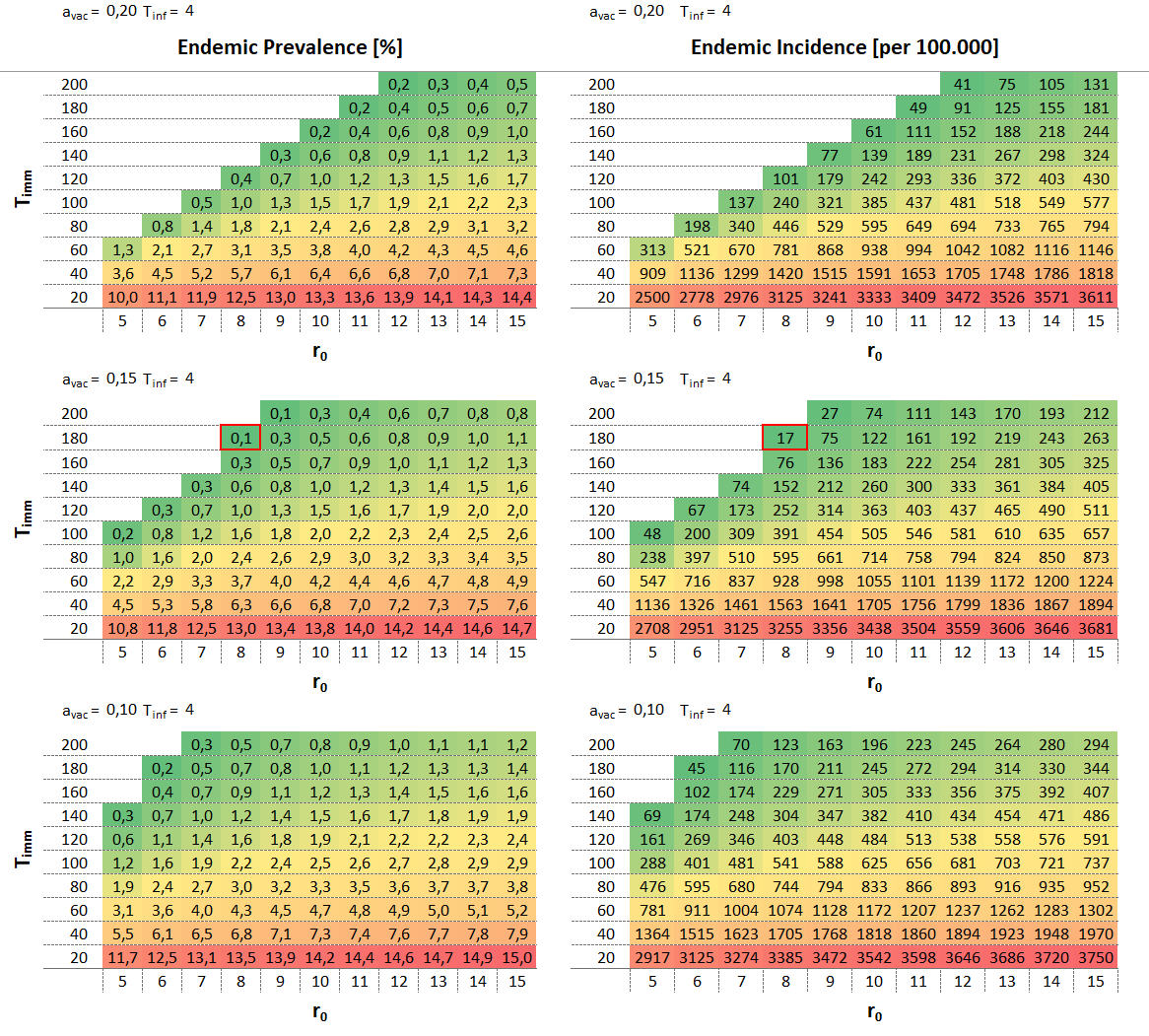}
	\caption{Prevalence and incidence tables at the endemic
	equilibrium for given values of the basic reproduction
	number $r_0$, the mean time of immunity $\Timm$ and the
	vaccination activity $\rv$.
	Time scales in days are fixed by assuming the mean
	time of infectiousness $\Tinf=4\,$days. }
	\label{Fig_prevalence}
\end{figure}

\begin{figure}[htbp!]
\centering
\includegraphics[width=0.9\textwidth]
	{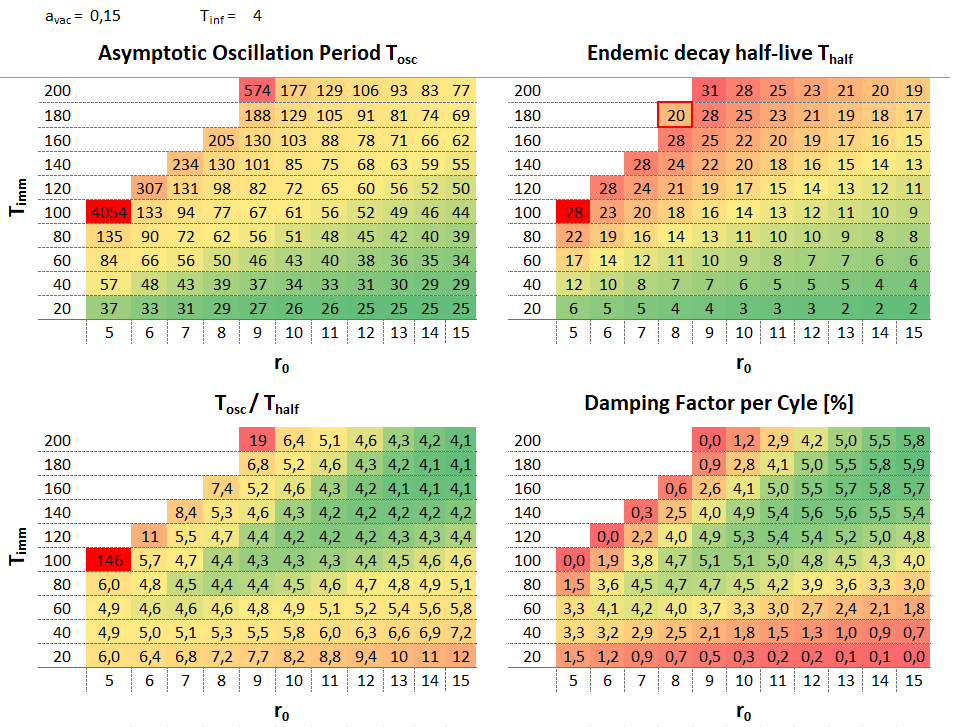}
	\caption{Asymptotic oscillation periods $\Tosc$ and decay
	half-lifes $\Thalf$ at fixed vaccination activity 
	$\rv=0.15$.
	The damping factor gives the asymptotic
	decay of oscillation amplitudes after one wave cycle.
	Time scales in days are fixed by assuming the mean
	time of infectiousness $\Tinf=4\,$days. }
	\label{Fig_time_scales}
\end{figure}

\bsn
I have then applied the SIRS model without vital dynamics to draw conclusions from latest data for the SARS-CoV-2 Omicron variant. In view of actual estimates  $r_{0,\mathrm{omikron}}\approx 7-14$ 
already this simplified model explains why the dynamics of  Omicron will most likely spiral into an endemic equilibrium. Vaccination programs are capable to reduce the final prevalence $I_2^*/N$ but are unlikely to prevent us from the oscillating scenario or even reach a disease free equilibrium. Yet, for a wide range of parameter values these oscillation effects would be very weak (see damping factors in Fig. \ref{Fig_time_scales}) and empirically presumably not be distinguishable from the non-oscillating endemic scenarios (C1)-(C2). Tables for endemic prevalence and incidence values predicted by this model are given in Fig \ref{Fig_prevalence}.\footnote{Of course one should be aware of under-reporting factors when comparing these values with officially reported numbers. For Germany these factors have lately been estimated between four and five in the first half of 2020 and reduced to roughly two starting with fall 2020, see the RKI-report from Aug. 2021 \autocite{RKI_underreporting}. Estimates for other countries partly seem to be much larger, for a systematic meta-analysis of 968 international studies with 9.3 Million probands from 76 countries see \autocite{Bobrovitz_et_al}.}

\bsn
Of course in many respects this model is too simple to describe reality quantitatively. In reality one has to face different behaviors of virus variants, vaccines, age groups, symptomatic severities and immunity responses by vaccination vs. recovery. Also incubation times are not negligible and estimates for 
the time of infectiousness are overruled by quarantine measures and hospitalization rates. But most importantly, the effective contact rate $\beta$ and hence $r_0$ are time varying due to seasonal effects, contact behaviors  and regional authority measures. So from this argument alone ongoing seasonal infection waves will completely overrule the weak endemic oscillations predicted by the autonomous SIRS model.

\appendix
\section{SIRS Tables at Endemic Equilibria\label{Sec_SIRS-tables}} 
This Appendix depicts some tables  of values predicted by the SIRS model at  endemic equilibria. Shown are prevalence and incidence values as well as oscillation periods and decay half-lives. Parameter ranges are 
$r_0\in[5,15]$, $\Timm/\Tinf\equiv a\inv\in [5,50]$ and 
$\rv\in\{0.10,\,0.15,\,0.20\}$. Truly time scales should be interpreted in units of $\Tinf\equiv\ga\inv$. To produce absolute numbers in days I have chosen 
$\Tinf=4\,$days throughout. For other choices of 
$\Tinf$ time scales would have to be rescaled accordingly.
The endemic prevalence $I_2^*/N$ is obtained from \Eqref{SI_2*} and the incidence at the endemic equilibrium is given by $\ga I_2^*/N$.
Formulas for the oscillation period $\Tosc$ and the decay-half time $\Thalf$ in scenario (C3) have been given in Eqs. \ref{Tosc} and \ref{Thalf}. 

In Figs. \ref{Fig_prevalence}  and \ref{Fig_time_scales} white cells fall into the disease-free and colored cells into the 	spiral endemic equilibrium. An exception is the cell 	
$r_0=8$, $\Timm=180$ and $\rv=0.15$ (read border), 	which belongs to the non-oscillating endemic scenario (C1).
In all other border cells scenario (C1) doesn't appear since parameter ranges for this scenario are too narrow to show up in the chosen resolution. 

\bsn
\section{The vaccination-reduced reproduction number\label{Sec_threshold} }
In models with more than one infectious compartment the notion of basic reproduction number has to be refined. 
\autocite{Diekmann_et_al}, see also \autocite{DiekmannHeesterbeek}, have defined a generalized reproduction number $\R_0$ given by the spectral radius of the next generation matrix. Using this definition and quite general axioms for  compartmental epidemic models \autocite{Driesche_Watmough2002, Driesche_Watmough2008} have shown that for 
$\R_0 < 1$ the disease-free equilibrium is locally asymptotically stable and for $\R_0>1$ it becomes 
unstable.\footnote{For sufficient conditions guaranteeing global stability for $\R_0<1$ see e.g. \autocite{Castillo-Chavez_et_al, Driesche_Watmough2008} or more recently
\autocite{AvramAdenane2022}.}
Moreover, in the case of just one infectious compartment, 
$\R_0$ coincides with the replacement number at the disease-free equilibrium. In our case, by looking at \Eqref{xy_1*}, this gives
\begin{equation}
\R_0=x_1^*=r_0c/b=r_0\frac{\al+\delta}{\sigma+\al+\delta}\,.
\end{equation}
Hence \Eqref{scenarios} verifies the above result, i.e. scenario (C) corresponds to $\R_0\equiv r_0c/b>1$.
Also, if we switch off the vaccination term, $\si=0$, then $\R_0=r_0$. This is why in SIRS/SIS models  $\R_0$ is often called the {\em vaccination-reduced reproduction number}. Finally, using \Eqref{xy_2*} the formula for the endemic prevalence can now be rewritten as
\begin{equation}
I_2^*/N =(1-\R_0\inv)\frac{c}{1+a}
=(1-\R_0\inv)\frac{\al+\delta}{\ga+\al+\delta}\,,
\end{equation}
which generalizes the formula in \Eqref{SI_2*} to the case 
$\delta>0$.

\newpage
\printbibliography

\end{document}